\documentclass{article}
\newcommand{\argmax}{\operatornamewithlimits{argmax}}
\usepackage{spconf,amsmath,amssymb,graphicx,mathtools, nccmath}
\usepackage{scalerel}
\usepackage{caption,subcaption}
\usepackage{multirow}
\usepackage{float}
\usepackage{lipsum}
\usepackage{hyperref}
\usepackage{enumitem}
\usepackage[nodisplayskipstretch]{setspace}

\usepackage{booktabs}
\usepackage{mathtools}
\usepackage{etoolbox}
\usepackage{xcolor}
\usepackage[capitalize]{cleveref}
\usepackage{blindtext}
\usepackage{eso-pic,rotating,graphicx}

\newcommand{\subalign}[1]{%
	\vcenter{%
		\Let@ \restore@math@cr \default@tag
		\baselineskip\fontdimen10 \scriptfont\tw@
		\advance\baselineskip\fontdimen12 \scriptfont\tw@
		\lineskip\thr@@\fontdimen8 \scriptfont\thr@@
		\lineskiplimit\lineskip
		\ialign{\hfil$\m@th\scriptstyle##$&$\m@th\scriptstyle{}##$\hfil\crcr
			#1\crcr
		}%
	}%
}

\def\L{{\cal L}}
\makeatletter
\renewcommand{\section}{\@startsection
	{section}%
	{1}%
	{}%
	{-0.7\baselineskip}%
	{0.1\baselineskip}%
	{}}%
\renewcommand{\subsection}{\@startsection
	{subsection}%
	{2}%
	{}%
	{-0.3\baselineskip}%
	{0.1\baselineskip}%
	{}}%
\renewcommand{\subsubsection}{\@startsection
	{subsubsection}%
	{3}%
	{}%
	{-0.1\baselineskip}%
	{0.1\baselineskip}%
	{}}%
\g@addto@macro\normalsize{%
	\setlength\abovedisplayskip{5pt plus 2pt minus 2pt}
	\setlength\belowdisplayskip{5pt plus 2pt minus 2pt}
	\setlength\abovedisplayshortskip{4pt plus 2pt minus 2pt}
	\setlength\belowdisplayshortskip{4pt plus 2pt minus 2pt}
}
\makeatother

\newcommand\numberthis{\addtocounter{equation}{1}\tag{\theequation}}
\copyrightnotice{978-1-6654-7189-3/22/\$31.00~\copyright2023 IEEE}
\AddToShipoutPicture{\put(590,750){\rotatebox{270}{\scalebox{0.53}{Copyright 2023 IEEE. Published in the 2022 IEEE Spoken Language Technology Workshop (SLT) (SLT 2022), scheduled for 19-22 January 2023 in Doha, Qatar. Personal use of this material is permitted. However, permission for advertising or promotional purposes or  for creating new collective works  for resale or}}}}
\AddToShipoutPicture{\put(585,750){\rotatebox{270}{\scalebox{0.53}{redistribution to servers or lists, or to reuse any copyrighted component of this work  in other works, must be obtained from the IEEE. Contact: Manager, Copyrights and Permissions / IEEE Service Center / 445 Hoes Lane / P.O. Box 1331 / Piscataway, NJ 08855-1331, USA. Telephone: + Intl. 908-562-3966.}}}}

\title{HMM vs. CTC for Automatic Speech Recognition: \\Comparison Based on Full-Sum Training from Scratch}
\name{Tina Raissi $^{\star1}$, \thanks{$\star$ Denotes equal contribution}Wei Zhou $^{\star 1,2}$, Simon Berger $^{1,2}$, Ralf Schl\"uter $^{1,2}$, Hermann Ney $^{1,2}$ \vspace{-2mm} }

\address{
	$^1$Human Language Technology and Pattern Recognition Group, RWTH Aachen University, Germany \\ $^2$AppTek GmbH, Aachen, Germany \\  \vspace{-2mm}
	 \textit{\{raissi,zhou,sberger,schlueter,ney\}@cs.rwth-aachen.de}}

\begin{document}
	
	\maketitle
	\begin{abstract}\vspace{1mm}
		In this work, we compare from-scratch sequence-level cross-entropy (full-sum) training of Hidden Markov Model (HMM) and Connectionist Temporal Classification (CTC) topologies for automatic speech recognition (ASR).\ Besides accuracy, we further analyze their capability for generating high-quality time alignment between the speech signal and the transcription, which can be crucial for many subsequent applications.\ Moreover, we propose several methods to improve convergence of from-scratch full-sum training by addressing the alignment modeling issue.\ Systematic comparison is conducted on both Switchboard and LibriSpeech corpora across CTC, posterior HMM with and w/o transition probabilities, and standard hybrid HMM.\ We also provide a detailed analysis of both Viterbi forced-alignment and Baum-Welch full-sum occupation probabilities.
		
	\end{abstract}
	\section{Introduction}
	\label{sec:intro}
	The recent sequence-to-sequence (seq2seq) acoustic models allow for from-scratch training within a unified optimization framework for automatic speech recognition~(ASR).\ The common sequence-level cross-entropy training for both transducer based models with different label topologies~\cite{ctc,rnnt,rnaSak}, and attention based encoder-decoder model~\cite{chorowski2015attention} do not necessarily require an initial alignment between the speech signal features and the output labels.\ The simplest case among the mentioned approaches is Connectionist Temporal Classification (CTC).\ The presence of the blank in the CTC topology gives more freedom to the alignment model which has less convergence problems during full-sum training, demanding at the same time less computation due to the independence assumption at each time step.\ However, the peaky behavior caused by the limited label emissions can also affect the quality of the alignment, which can shift with respect to the evidence in the input.\ 
	
	The typical training pipelines deployed for standard hybrid hidden Markov model deep neural network~(HMM-DNN)~\cite{bourlard1994hybrid} on the other hand requires the bootstrapping of a separate context-dependent Gaussian Mixture Model~(GMM).\ The alignment taken from the GMM system is generally known to be a good starting point for subsequent ASR tasks or data segmentation, despite its low ASR accuracy.\ The hybrid model is generally trained with frame-wise cross-entropy~(Fw-CE) using GMM alignment with a multi-stage complex pipeline consisting of various training criteria.\
	
	Previous work on single stage from-scratch systems considered Lattice-Free Maximum Mutual Information criterion~(LF-MMI)~\cite{mmi,endtoendpovey} and its further extension~\cite{boostmmi}, using both CTC and HMM.\ However, the training can require high computations with still not fully competitive ASR results.\ Another possible from-scratch training solution is to use the maximum likelihood~(ML) criterion, which is shown to have convergence problems that require complex training schedules and even the inclusion of additional losses~\cite{alberteugen}.\ Moreover, the quality of the Viterbi alignment taken from such from-scratch ML trained HMM based model even with an extended multi-task loss is still shown to lead to an inferior performance compared to a GMM based alignment, when used for further Fw-CE training in hybrid approach~\cite{tinatowards}.\
	
	In this work, we compare the CTC and HMM topologies in the context of from-scratch full-sum training.\ In addition to the standard hybrid HMM, we propose a discriminative HMM modeling as a more direct contrast to CTC models.\ We then address the convergence issue from point of view of difficulty of alignment modeling, and propose several approaches in that regard.\ Together with the common subsampling, we propose both minimum label duration and prior-knowledge-based probability approximation.\ Experiments on both Switchboard and LibriSpeech copora show that these approaches not only benefit HMM models, but also improve CTC models.\ Besides ASR accuracy, we also analyze their capability of generating good alignment.\ With all these, we hope to initiate the path towards a simple from-scratch trainable system producing both good word error rate and alignment.

	\section{Models}
	\label{sec:model}
	Let X and W denote the acoustic feature sequence and corresponding word sequence of a speech utterance, respectively. 
	Let $h_1^T = E(X)$ denote the common encoder output which transforms $X$ into high-level representations with optional subsampling.\ In the following, we consider the two most common label topologies for time-synchronous 0-order models, namely CTC and HMM, in the context of from-scratch full-sum training.
	With a major focus on phonetic-based acoustic modeling (AM), we provide detailed formulations in terms of models, training and decoding.\
	We denote $a_1^S$ (usually $T > S$) as the phonetic output label sequence of $W$ and omit pronunciation variants for simplicity.\
	
	\begin{figure}[t]
		\centering
		\includegraphics[width=0.8\linewidth]{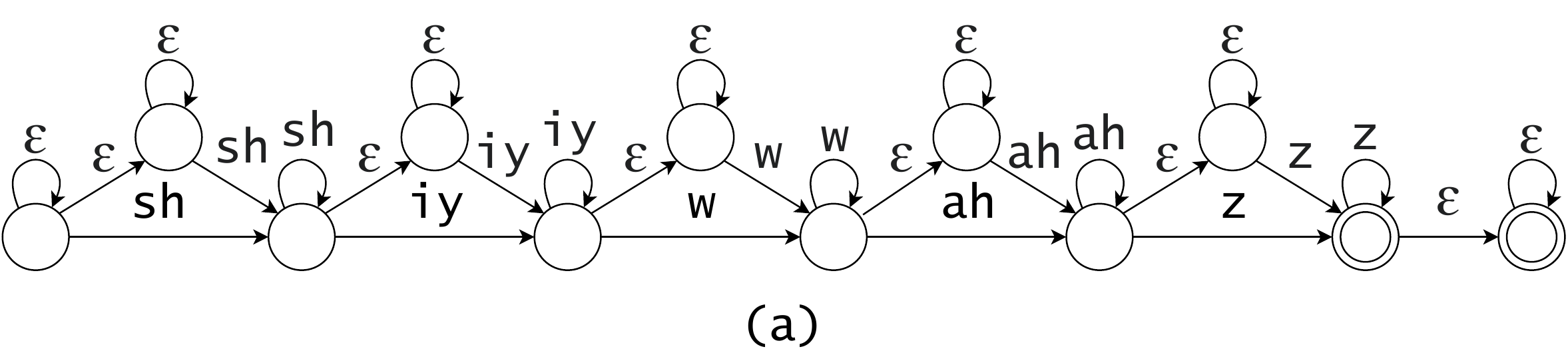}
		\includegraphics[width=0.8\linewidth]{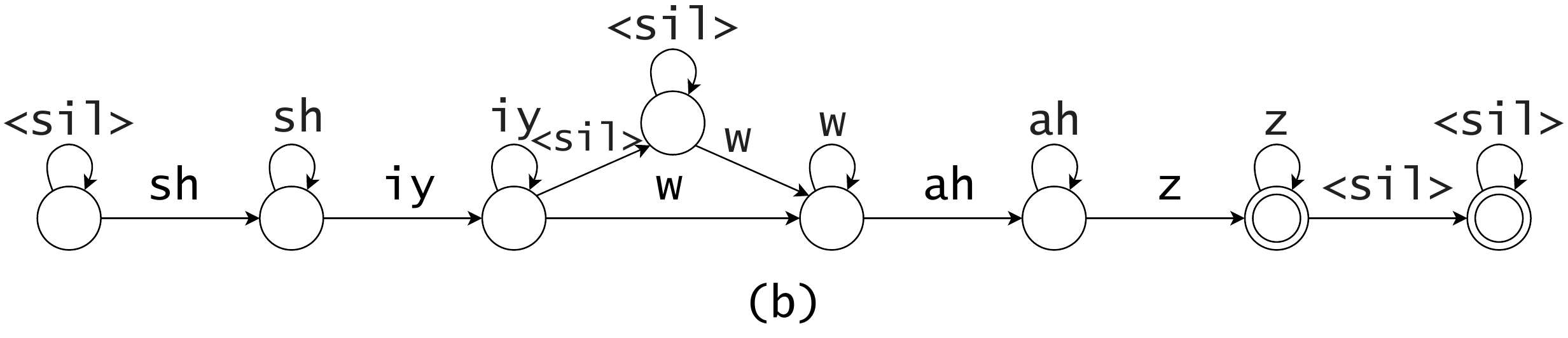}
		\caption{FSA example of (a) CTC and (b) HMM topologies for the phoneme sequence of the utterance ``\textit{she was}"}
		\label{fig:fsa}
	\end{figure}
	\ClearShipoutPicture
	
	\subsection{CTC}
	\label{subsec:ctc}
	The standard CTC topology~\cite{ctc} introduces blank $\epsilon$-augmented alignment sequences $y_1^T$ to align $a_1^S$ to the $T$ frames.
	An example of possible alignments in the form of finite state automaton (FSA) is shown in \cref{fig:fsa}.
	Each $a_s$ has to occur at least once in $y_1^T$ and optional $\epsilon$ can occur at any label segment boundaries. 
	For label repetitions in $a_1^S$, at least one $\epsilon$ has to occur between the two labels in $y_1^T$.
	This allows a unique mapping of $y_1^T$ to $a_1^S$ by removing all label loops and $\epsilon$. 
	Thus, the posterior of the output label sequence can be written as: 
	\begin{equation}
		P(a_1^S | X) = \sum_{y_1^T:a_1^S} P(y_1^T | h_1^T) = \sum_{y_1^T:a_1^S} \prod_{t=1}^T P(y_t | h_t)  \label{eq:ctc}
	\end{equation}
	Based on \cref{eq:ctc}, the full-sum training for CTC models can be carried out from-scratch with the following loss:
	\begin{equation}
		\L = -\log \hspace{1mm} P(a_1^S | X) \label{eq:fsloss}
	\end{equation}
	The final best word sequence $\hat{W}$ can be obtained via Viterbi decoding as:
	\begin{equation} 
		X \rightarrow \hat{W} = \underset{W}{\argmax} \hspace{1mm} \Big[ P^{\lambda}_{\text{LM}}(W) \underset{y_1^T:a_1^S:W}{\max} \prod_{t=1}^T \frac{P(y_t | h_t)}{P^{\alpha}_{\text{prior}}(y_t)} \Big] \label{eq:ctcdecode}
	\end{equation}
	Here an optional $P_{\text{prior}}$ with scale $\alpha$ can be included, which can be obtained by marginalizing the framewise posterior on the training data.
	For $\alpha=0$, this corresponds to a simple log-linear combination of the AM and language model~(LM) with scale $\lambda$.

	\begin{figure}[t]
		\centering
		\includegraphics[width=0.8\linewidth]{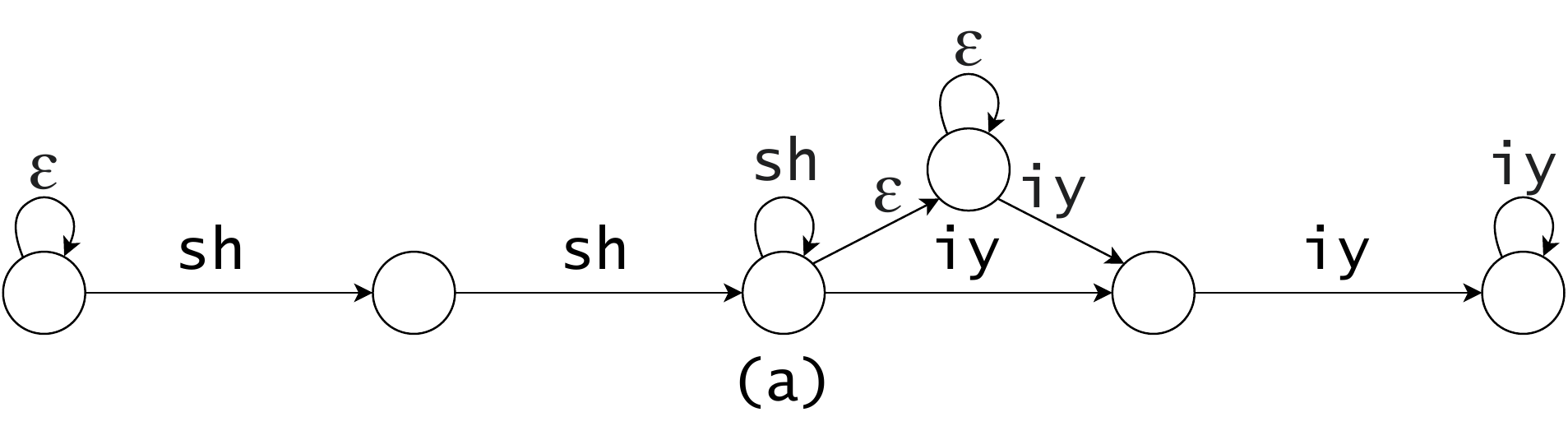}
		\includegraphics[width=0.8\linewidth]{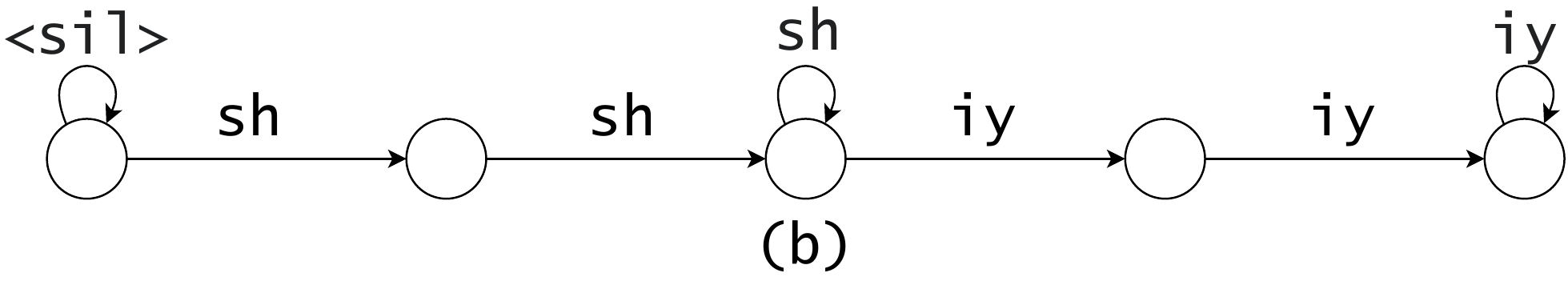} 
		\caption{Modified FSA of (a) CTC and (b) HMM topologies with MinDur (k=2) for the first word ``\textit{she}" in \cref{fig:fsa}}
		\label{fig:minDur}
		
	\end{figure}
	\subsection{HMM}
	\label{subsec:hmm}
	In this work, we mainly consider the HMM-0-1 topology, where only forward and loop transitions are allowed.
	As shown by the FSA	in \cref{fig:fsa}, the alignment sequence in this case contains only $a_s$ with optional loops, while optional silence is also introduced at word boundaries.
	To be consistent with standard HMM notations, here we use the state sequence $s_1^T$ to represent alignments for $a_1^S$ as $y_t = a_{s_t}$, which is also necessary to distinguish label repetitions in $a_1^S$.
	Note that theoretically there is no deterministic bijective mapping between the HMM and CTC topologies for any alignments.
	
	We study the HMM topology with both discriminative and generative modeling approaches. We adopt the Markov assumption for state dependency and further simplify that with a pooled transition $\mathcal{T}(s_t, s_{t-1})$, which only has 4 values for loop/forward of speech/silence.

	\subsubsection{Posterior HMM}
	\label{subsubsec:p-hmm}
	We call the discriminative approach as posterior HMM (\textbf{P-HMM}), where the output sequence posterior is defined as:
	\begin{align*}
		P(a_1^S| X) &= \underset{s_1^T}{\sum} P(a_1^S, s_1^T |h_1^T) \\
		&= \underset{s_1^T:a_1^S}{\sum} \prod_{t=1}^T P(a_{s_t}, s_t | s_{t-1}, h_t) \\
		&= \underset{s_1^T:a_1^S}{\sum} \prod_{t=1}^T P(a_{s_t}|s_{t-1}^t, h_t) P(s_t| s_{t-1}, h_t) \\
		&= \underset{s_1^T:a_1^S}{\sum} \prod_{t=1}^T P(a_{s_t}|h_t) \mathcal{T}(s_t, s_{t-1})\numberthis \label{eq:phmm}
	\end{align*}
	The full-sum training of P-HMM can then be carried out from scratch based on \cref{eq:fsloss} and \cref{eq:phmm}, where we can further apply exponential scales on the two probability terms in \cref{eq:phmm}.
	
	Similar as \cref{eq:ctcdecode}, the final best word sequence $\hat{W}$ can be obtained via Viterbi decoding as:
	\begin{align*}	 
		&X \rightarrow \hat{W} =  \numberthis \label{eq:phmmdecode} \\ 
		&\underset{W}{\argmax} \hspace{1mm} \Big[ P^{\lambda}_{\text{LM}}(W) \underset{s_1^T:a_1^S:W}{\max} \prod_{t=1}^T \frac{P(a_{s_t} | h_t)}{P^{\alpha}_{\text{prior}}(a_{s_t})} \mathcal{T}^{\beta}(s_t, s_{t-1}) \Big]  
	\end{align*} 
	where the optional $P_{\text{prior}}$ can still be applied.
	
	To explore the strong similarity with CTC models, we further simplify the P-HMM with a fixed $\mathcal{T}(s_t, s_{t-1})=0.5$ for all types of transitions. 
	This allows to directly drop the term $\mathcal{T}(s_t, s_{t-1})$ in both training and decoding, which yields the correlation between \cref{eq:ctc} vs. \cref{eq:phmm} and \cref{eq:ctcdecode} vs. \cref{eq:phmmdecode}.
	We call this simplified version as \textbf{P-HMM-S}.  
	\subsubsection{Hybrid HMM}
	\label{subsubsec:h-hmm}
	The generative modeling approach is the classical hybrid HMM (\textbf{H-HMM}), which defines
	\begin{align*}
		P(X| a_1^S) &= P(h_1^T | a_1^S) = \underset{s_1^T}{\sum} P(h_1^T, s_1^T | a_1^S) \\
	   &= \underset{s_1^T:a_1^S}{\sum} \prod_{t=1}^T P(h_t | a_{s_t}) P(s_t | s_{t-1}) \\
		&= \underset{s_1^T:a_1^S}{\sum} \prod_{t=1}^T \frac{P(a_{s_t} | h_t)}{P_{\text{prior}}(a_{s_t})} \mathcal{T}(s_t, s_{t-1}) \numberthis \label{eq:hhmm}
	\end{align*}
	The from-scratch full-sum training of H-HMM can then be carried out by minimizing the loss
	\begin{align*}
		\L = -\log \hspace{1mm} P(X | a_1^S)\numberthis
		\label{eq:losshhmm}
	\end{align*}
	where we can further apply exponential scales on the three probability terms in \cref{eq:hhmm}. Despite the tuning effort, we find that an optimal scaling is essential for both convergence and final best performance.
	
	The decoding of the best word sequence using H-HMM is the same as \cref{eq:phmmdecode}.
	Note that the major difference to P-HMM is the inclusion of $P_{\text{prior}}$ in training, which effectively results from the discriminative and generative modeling nature.
	\subsection{Label Units}
	\label{subsec:lu}
	Following~\cite{wei}, we apply end-of-word (EOW) augmentation to the phoneme sets by default. For the granularity of $a_1^S$, we consider both single-state and three-states structure for each phoneme. The latter is commonly applied in the classical hybrid HMM approach to account for the start, middle, and end of a phoneme.
	
	\section{Convergence and Alignment Modeling}
	\label{subsec:align}
	The full-sum training with either CTC or HMM topology is effectively an Expectation Maximization~(EM)-style alignment modeling problem.\ The general procedure relies on the calculation of the occupation probability $q_t$, known as soft-alignment, at each EM iteration.\ The computation of the soft-alignment can be done via the Baum-Welch forward-backward algorithm.\ For a given input $h_1^T$, and a label sequence $a_1^S$, the quantity $q_t(a_s | h_1^T, a_1^S, \theta)$ represents the probability mass of all alignment paths going through the state $s$ at time frame $t$, according to the parameters $\theta$.\ In practice, the quantity $q_t$ is a normalized probability distribution over the set of all labels at time frame $t$.\ With the maximum approximation, the label of aligned state in the best path has probability one.\ The latter alignment is known as hard-alignment or Viterbi forced alignment.\ 
	
	Ideally, the model in training should converge to more prominent alignment paths and assign higher probabilities to them.\ The difficulty of such alignment modeling can be affected by the number of possible alignment paths, which is correlated with the relation between $T$ and $S$ (usually $T \gg S$). This is then directly related to the convergence issue of neural network (NN) training (generally non-convex optimization), especially from scratch.
	
	With blank allowed almost anywhere, the CTC topology reveals an easy optimum for NN training to assign most probability masses to alignments with most blank frames.\ This allows an easy convergence for CTC models, but also leads to the common peaky behavior~\cite{zeyer2021:peakyctc}.\ Thus, CTC models usually is not able to produce Viterbi alignment with high correspondence to the actual speech signal, where the ambiguity of blank even enlarges the difficulty~\cite{boostmmi}.\ To solve this, one might need to force certain label loops and force blank to cover silence only, which eventually forces the CTC topology towards the HMM topology.\ On the other hand, the HMM topology appears to be more restricted and thus, may result in more difficulty in convergence~\cite{alberteugen}.
	Here we propose to solve the convergence issue by reducing the difficulty of alignment modeling:  \vspace{-2mm}

	\begin{itemize}[leftmargin=*, itemsep=-1.2mm]
		\item \textbf{subsampling (SS)} is widely applied in modern ASR systems to reduce computation and memory cost. By reducing T, subsampling effectively also reduce the number of possible alignment paths and therefore, can largely simplify alignment modeling for an easier convergence (extreme case: $T \approx S$).
		\item \textbf{minimum duration (MinDur)} of speech label forces each $a_s$ to occur at least $k$ times in the alignment. This shares the same spirit as subsampling to reduce the number of possible alignment paths and thus, can also ease the convergence issue.
		In addition, it also forces the model to be less peaky, which might lead to more reasonable alignment paths.
		The minimum duration, as shown in \cref{fig:minDur}, can be simply applied by modifying the FSA structure for full-sum training.
		\item \textbf{probability approximation ($P$-approx)} uses prior knowledge to estimate those transition and prior probabilities in P-HMM or H-HMM. These values are then kept fixed during training to guide the model towards more reasonable alignment paths.
		For our phonetic-based AM, we make use of the fact that a phoneme is commonly of 80ms long on average, which reveals the speech forward/loop probabilities as $\mathcal{T}(s_t=s_{t-1})=7/8$ under a 10ms-frame scenario.
		Together with the transcription of acoustic training data, we can also derive the prior probability of each phoneme.
		Then by regarding the remaining audio length of each utterance all to be silence, we can also obtain the silence prior probability.
		Finally, by regarding these silence length of each utterance as equally from sentence begin/end, we obtain the silence transition probability for $\mathcal{T}$.
	\end{itemize}
     These methods can also be combined with minor adjustment. We also apply subsampling and minimum duration to CTC models for fair comparison.
	\section{Experiments}
	\label{sec:exp}
	\subsection{Setting}
	\label{subsec:set}
	
	We conduct our experiments on two separate corpora, 960 hours LibriSpeech~(LBS) \cite{panayotov2015librispeech}, and 300 hours Switchboard (SWB) (LDC97S62)~\cite{godfrey1992switchboard}.\ The evaluations are done on all four dev and test sets for LBS, as well as on SWB and CallHome~(CH) subsets of Hub5'00 (LDC2002S09) and the three SWB subsets of Hub5'01 (LDC2002S13). 
	We use the official lexicon of LBS and SWB, where we unify the stressed phonemes for the former~\cite{weiefficient}.
	EOW phonemes are applied in all cases.

	\begin{table}[t]
		
		\setlength{\tabcolsep}{0.2em}\renewcommand{\arraystretch}{.95}  
		\centering \footnotesize 
		\caption{Performance of our models in terms of word error rates (WER) on Hub5'0\{0,1\} using a 4-gram language model.\  We show the effect of using SS, MinDur and $P$-approx (for $\mathcal{T}$ and/or $P_{\text{prior}}$).\ We consider single-state or three-states phoneme with EOW.\ We denote $\times$ as not converged model. }
		\label{tab:SWB-WER}
		\begin{tabular}{|c|c||ccc||c|c|} 
			\hline 
			\multirow{2}{*}{\textbf{Model}} & \multirow{2}{*}{ \textbf{States} } &  \multirow{2}{*}{ \textbf{SS} } &  \multirow{2}{*}{\textbf{MinDur} } & \multirow{2}{*}{\textbf{$\mathbf{P}$-approx}} &\multicolumn{2}{c|}{ \textbf{WER [\%]}} \\ \cline{6-7}
			& 	 &   &    									&      &  {Hub5'00}& {Hub5'01} \\ \hline
			{\scriptsize \multirow{3}{*}{\textbf{CTC}}  }  & \multirow{8}{*}{1} &   &  &  & $17.6$& $16.0$ \\ \cline{3-7}			
			
			 & 	           & \checkmark  &          & & $\mathbf{13.7}$ & $\mathbf{13.1}$  \\ \cline{3-7}			
			&          &              &   \checkmark & & $14.6$& $14.0$\\  \cline{1-1} \cline{3-7}			
			
			{\scriptsize \multirow{3}{*}{\textbf{P-HMM-S}} }&          &	   &  	     &          &   \multicolumn{2}{c|}{$\times$} \\ \cline{3-7}			
			&               & \checkmark  &           & &  $14.5$ & $13.9$ \\ \cline{3-7}			
			&               &         & \checkmark   &   &  $15.0$&   \multirow{2}{*}{-} \\   \cline{1-1}\cline{3-6}			
			
			{\scriptsize \multirow{3}{*}{\textbf{P-HMM}}}     &     &    		&        &    \checkmark       & $17.5$ &  \\   \cline{3-7} 			
			&      &     \checkmark      &  &   \checkmark& $14.3$ & $13.9$ \\   \cline{2-7}	
			&  3    &          &  &   \checkmark& $16.2$ &   \multirow{2}{*}{-} \\ \cline{1-6}
			{ \scriptsize \multirow{2}{*}{\textbf{H-HMM}}  }  &    1   &  &   &  \checkmark & $24.0$ &  \\ \cline{2-2}\cline{3-7}

			& 	 3 &      &   & \checkmark  &  $13.9$ &$13.5$  \\  \hline
			
		\end{tabular} 
	\vspace{-0.5mm}
	\end{table}
	
	\begin{table}[t]
		\setlength{\tabcolsep}{0.1em}\renewcommand{\arraystretch}{.95}  
		\centering \footnotesize
		\caption{Similar experiments for the most promising models presented in \cref{tab:SWB-WER}, trained with LibriSpeech 960 hours, and evaluated on all four test and dev datasets.}
		\label{tab:LBS-WER}
		\begin{tabular}{|c|c||ccc||c|c|c|c|} 
			\hline
			
			& 	 &   &    									   &   &\multicolumn{4}{c|}{ \textbf{WER [\%]}} \\ \cline{6-9}
			\multirow{3}{*}{\textbf{Model}}    & \multirow{3}{*}{ \textbf{States} } & \multirow{3}{*}{ \textbf{SS} } &  \multirow{3}{*}{\textbf{MinDur} } & \multirow{3}{*}{\textbf{$\mathbf{P}$-approx}}& \multicolumn{2}{c|}{ \textbf{dev}}&  \multicolumn{2}{c|}{ \textbf{test}}  \\ \cline{6-9}
			& 	 &   &    									    &   &  clean& other& clean & other \\ \hline
			
			\multirow{3}{*}{\textbf{CTC}}      & \multirow{7}{*}{1}     & & &  &$4.6$& $11.3$  & $5.0$ & $12.3$ \\ \cline{3-9}	
			& 	           & \checkmark  &       &  & $\mathbf{3.2}$& $\mathbf{8.0}$ & $\mathbf{3.6}$& $\mathbf{8.5}$\\   \cline{3-9}	
			& 	           &   &    \checkmark    & & $3.7$& $8.9$ & $4.1$ & $9.6$  \\  \cline{1-1} \cline{3-9}	
			
			{\scriptsize \multirow{3}{*}{\textbf{P-HMM-S}} }           &&&&	   &  \multicolumn{4}{c|}{ $\times$ }	     \\ \cline{3-9}
			&         	   &  	\checkmark     &          &  & $3.4$ & $8.5$& $3.8$ & $9.4$ \\ \cline{3-9}	
			&          &              & \checkmark&   & $3.7$& $9.1$&  \multicolumn{2}{c|}{\multirow{2}{*}{-}} \\ \cline{1-1} \cline{3-7}	
			{\scriptsize \multirow{1}{*}{\textbf{P-HMM}}  }   &     &\checkmark	&         &      \checkmark     & $3.5$& $8.9 $&  \multicolumn{2}{c|}{} \\  \hline
			{\scriptsize \multirow{1}{*}{\textbf{H-HMM}}  } & 	 3  &   & & \checkmark  & $3.6$ & $9.1$& $3.9$& $9.5$ \\  \hline
			
		\end{tabular}  
	\vspace{-2mm}
	\end{table}
	
\begin{table}[t]	
	\setlength{\tabcolsep}{0.07em}\renewcommand{\arraystretch}{.95}  
	\centering \footnotesize
	\caption{Evaluation of our best model for SWB together with some results from the literature.\ WE denote monophone and diphone  as \{mono,di\}P, and specify the number of states (\{1,3\}-S), with optional EOW augmentation.\ }
	
	\label{tab:lit-SWB}
		\begin{tabular}{|c|c|c|c|c|c||c|c|c|} 
			\hline			
			\multirow{2}{*}{\textbf{Work}} &\multirow{2}{*}{\textbf{Model}} & \multirow{2}{*}{\textbf{Label}}&  \multirow{1}{*}{\textbf{From-}}& \multirow{2}{*}{\textbf{Criterion}} &\multirow{2}{*}{\textbf{LM}} &\multicolumn{3}{c|}{\textbf{WER [\%]} }  \\ \cline{7-9}
			& & & \textbf{Scratch} & & & \textbf{SWB} & \textbf{CH}  &\textbf{$\sum$} \\ \hline \hline
			\cite{tinafh} &  H-HMM    & monoP-3S & no  &Fw-CE  &    4-gram   & $9.8$ & $\mathbf{18.4}$  & $14.1$ \\ \hline \hline
			\cite{hadian2018flat} & HMM  &  diP-2S  &   \multirow{3}{*}{yes} & LF-MMI & RNN  &    $9.8$    & $19.3$ & \multirow{2}{*}{ - }\\  \cline{1-3} \cline{5-8}
			\cite{ibmctc} &   \multirow{2}{*}{ CTC }  & 1P-1S &  &   \multirow{3}{*}{ML}  & \multirow{3}{*}{4-gram}   &    $13.9$    &  $24.7$   &  \\ \cline{1-1} \cline{3-3} \cline{7-9}			  
			This &   & EOW-1P-1S & &  &      & $\mathbf{8.8}$  & $18.5$  & $\mathbf{13.7}$  \\\cline{7-9} \cline{2-3}
			Work		&  H-HMM  &  EOW-1P-3S &  &   &    &$9.3$ & $18.5$ & $13.9$ \\   \hline			
		\end{tabular}
	\end{table}	
	\begin{table}[t]
		
		\setlength{\tabcolsep}{0.07em}\renewcommand{\arraystretch}{.95}  
		\centering \footnotesize
		\caption{Comparison of our models with other approaches in the literature for LBS 960h, with evaluation on all dev and test sets.\ We considered also triphone CART label~(triP) and wordpieces~(WP).}
		\label{tab:lit-LBS}
		\begin{tabular}{|c|c|c|c|c|c||c|c|c|c|} 
			\hline			
			\multirow{2}{*}{\textbf{Work}} &\multirow{2}{*}{\textbf{Model}} & \multirow{2}{*}{\textbf{Label}}&  \multirow{2}{*}{\textbf{From-}}& \multirow{2}{*}{\textbf{Criterion}} &\multirow{2}{*}{\textbf{LM}}   &\multicolumn{4}{c|}{ \textbf{WER [\%]}} \\ \cline{7-10}
			& 	 &   &    	 \multirow{2}{*}{\textbf{Scratch}}	&   & & \multicolumn{2}{c|}{ \textbf{dev}}&  \multicolumn{2}{c|}{ \textbf{test}}  \\\cline{7-10}
			& 	 &   &    									&   &   &  clean& other& clean & other \\ \hline\hline
			\cite{grazpaper} &  H-HMM    & triP-3S& no &Fw-CE  &   4-gram &$4.0$&$9.6$&$4.4$& $10.0$\\ \hline \hline
			\cite{boostmmi} & HMM & WP &  \multirow{4}{*}{yes} & bLF-MMI &  4-gram &  $3.9$ & $9.7$ &  \multicolumn{2}{c|}{-} \\ \cline{1-3}   \cline{5-10}			
			\cite{wave2letter} & Wav2L &   1P-1S&&   \multirow{3}{*}{ML}  & KenLM&  \multicolumn{2}{c|}{-} &$7.2$&- \\  \cline{1-3} \cline{6-10}
			This                      & CTC  &   EOW- &  &  &     \multirow{2}{*}{4-gram}    &$\mathbf{3.2}$  & $\mathbf{8.0}$  & $\mathbf{3.6}$ & $\mathbf{8.5}$ \\\cline{7-10} \cline{2-2}
			Work		           &  P-HMM-S  & 1P-1S &&&& 3.4 & 8.5 & 3.8 & 9.4 \\  \hline
		\end{tabular}
		\vspace{-4mm}
	\end{table}
	
	All acoustic models use a $6 \times 512$ BLSTM encoder and the optional subsampling of factor 4 is done via 2 max-pooling layers, in the middle.\ The input features to the encoder are (LBS: 50 and SWB: 40) dimensional Gammatone Filterbank features~\cite{schluter2007gammatone}, extracted from 25 milliseconds~(ms) window with 10ms shift.\ We train our models for (LBS: 25 and SWB: 30) epochs, combining two learning rate~(LR) scheduling.\ An initial one-cycle learning rate (OCLR) policy~\cite{smith2019super} covers $90\%$ of the total steps with linear increase up to peak LR of around $3e^{-3}$ and consequent linear decrease, following~\cite{weiefficient}.\ We continue the remaining steps with a constant minimum LR of $1e^{-5}$.\ The Adam optimizer with Nesterov momentum~\cite{dozat2016incorporating} is used.\  
	As regularization techniques we use $10\%$ dropout~\cite{srivastava2014dropout} and $L_2$ weight decay with a scale of $1e^{-4}$.\ Training of P-HMM and H-HMM includes also the choice of the scales $\alpha$, $\beta$, and $\gamma$ for the state prior, transition and label posterior, respectively.\
	The optimal setting for minimum duration is (LBS: 3, SWB: 4) and (LBS: 4, SWB: 5) for CTC and HMM, respectively.\
		
	Training of our models rely on the combination of RASR and RETURNN toolkits~\cite{wiesler2014rasr, zeyer2018returnn}, where the weighted FSA for each utterance is constructed within the former, and the NN training and the CUDA based BW computations are done by the latter.\ The decoding makes use of time synchronous lexical prefix trees search within RASR~\cite{ney1992improvements,ney1999dynamic}.
	By default we use word-level 4-gram LM in all cases.\

	\subsection{ASR Accuracy}
	\label{subsec:res}
	We present our results for different modeling approaches in \cref{tab:SWB-WER,tab:LBS-WER}, and compare our best model with the literature in \cref{tab:lit-SWB,tab:lit-LBS}.\ For fair comparison, we consider only from-scratch 0-order models from the literature, when possible.\
	
	We show that the application of either subsampling or minimum duration to CTC improves the accuracy on both corpora, with the former performing the best among all proposed models.\ Similarly, we can see that P-HMM-S with subsampling performs better than with minimum duration, on both corpora.\ The HMM topology without any additional method does not converge for any of the two copora.\ On SWB task we observe that by adding the transition probability the model converges to $17.5\%$ WER on Hub5'00.\ This can be improved by either switching to a three-states model or by application of subsampling, reaching $16.2\%$ and $14.3\%$, respectively.\ The combination of subsampling and fixed transition helped the P-HMM also on LBS task, obtaining $3.5\%$ and $8.9\%$ on dev-clean and dev-other, respectively.\ The H-HMM with its $13.9\%$ and $13.5\%$ on the two Hub5 datasets, show only relative $1\%$ and $2\%$ degradation compared to the CTC with subsampling.\ However, this gap becomes larger on LBS task. 
	
	Our best from-scratch models with CTC and HMM topologies on both SWB and LBS obtained competitive results compared to other approaches in the literature.\ Our H-HMM, as reported in \cref{tab:lit-SWB}, obtains slightly better performance than the hybrid monophone trained with a tandem alignment~\cite{zoltanalignment}.\ On LBS task included in \cref{tab:lit-LBS}, our P-HMM-S outperforms a triphone hybrid trained with triphone GMM alignment.\ CTC with subsampling again outperforms all models.\ 
		\begin{table}[t]
			
			\setlength{\tabcolsep}{0.3em}\renewcommand{\arraystretch}{.95}  
			\centering \footnotesize
			\caption{The calculation of TSE on SWB train and Hub5'00 dev-sub with respect to a GMM monophone alignment  for our proposed models and a standard BLSTM hybrid HMM trained with Fw-CE using a tandem based alignment.\ }
			\label{tab:tse}
			\begin{tabular}{|c|c|c||ccc||c|c|} 
				\hline			
				\multirow{2}{*}{\textbf{Plot}} &\multirow{2}{*}{\textbf{Model}}   &\multirow{2}{*}{\textbf{States}} & \multirow{2}{*}{\textbf{SS}}&  \multirow{2}{*}{\textbf{MinDur}}&\multirow{2}{*}{\textbf{$\mathbf{P}$-approx}} & \multicolumn{2}{c|}{\textbf{TSE [ms]}}  \\ \cline{7-8}
				& &    & 	 &   &    & train  &  dev-sub \\  \hline

				- &	{\scriptsize H-HMM-CE }  & 	3& &   &    & 41 & 39 \\ \hline
				a.1 & 	{\scriptsize\multirow{3}{*}{CTC} }  & \multirow{5}{*}{1} 	& &   &    & 86 & 67 \\ \cline{1-1} \cline{4-8}
				a.2 &  &   	 & \checkmark    &&  & 67 & 56 \\  \cline{1-1} \cline{4-8}
				a.3 &    & &	 &  \checkmark &   &  \textbf{52}& \textbf{49} \\  \cline{1-2} \cline{4-8}
				a.4 & 	{\scriptsize \multirow{2}{*}{P-HMM-S} } &	 & \checkmark  &   & & 68 & 56 \\  \cline{1-1}  \cline{4-8}
				a.5 &    	 &&    &  \checkmark&   & 72 & 68 \\  \cline{1-8}
				a.6 &    {\scriptsize P-HMM}	 &  \multirow{2}{*}{3}  && & \checkmark   & 63 &  57\\  \cline{1-2} \cline{4-8}
				a.7 & 	{\scriptsize	H-HMM } & &  &   & \checkmark & 106 & 89 \\  \hline
				
			\end{tabular}
	
		\end{table}
		
		\begin{table}[t]
			
			\setlength{\tabcolsep}{0.7em}\renewcommand{\arraystretch}{.95}  
			\centering \footnotesize
			\caption{The effect on convergence during training of the state prior, label posterior, and transition scales, $\alpha$, $\gamma$, and $\beta$, respectively.\ All experiments are for SWB300h and evaluated on Hub5'00 with 4-gram LM. }
			\label{tab:scales}
			\begin{tabular}{|c|c|c|c|c|c|} 
				\hline
				\textbf{Model}                                 & \textbf{Approach}                                           & \textbf{$\alpha$}         &  \textbf{$\gamma$}      &  \textbf{$\beta$}&  \textbf{WER [\%]} \\ \hline \hline
				\multirow{4}{*}{\textbf{P-HMM}}   &  \multirow{2}{*}{$P$-approx}           & \multirow{4}{*}{N/A} &  1.0                                 &  1.0 &  $\times$ \\  \cline{4-6}
				&  												                           &                                     &	0.3						      	    &  0.1 &  $17.5$ \\ \cline{2-2}  \cline{4-6}
				&   \multirow{2}{*}{$P$-approx+SS}   &						               & \multirow{2}{*}{1.0}	   &  0.1 &  $15.0$ \\  \cline{5-6}
				&  												                        	&									  &                                        &  0.01 &  $14.4$ \\  \hline \hline
				\multirow{5}{*}{\textbf{H-HMM}}      &  \multirow{5}{*}{$P$-approx}   &           1.0                   &  1.0                                 &  1.0 &  $\times$  \\  \cline{3-6}
				&  												                        	&				0.5					 &    \multirow{4}{*}{0.3}  &   \multirow{3}{*}{0.1} &  $20.3$ \\  \cline{3-3} \cline{6-6}
				&  												                        	&			0.3				         &                                        &  &  $17.5$ \\  \cline{3-3} \cline{6-6}
				&  												                        	&\multirow{2}{*}{0.1}  &           							        &   &  $15.1$ \\  \cline{5-6}
				&  												                        	&									  &                                         &  0.3&  $14.7$ \\  \hline
				
			\end{tabular}
		\vspace{-2mm}
		\end{table} 
		
		\begin{figure*}[t]
			\centering 
			\includegraphics[height=.75\linewidth, clip, trim=0.3cm 0.55cm 0.3cm 1.cm]{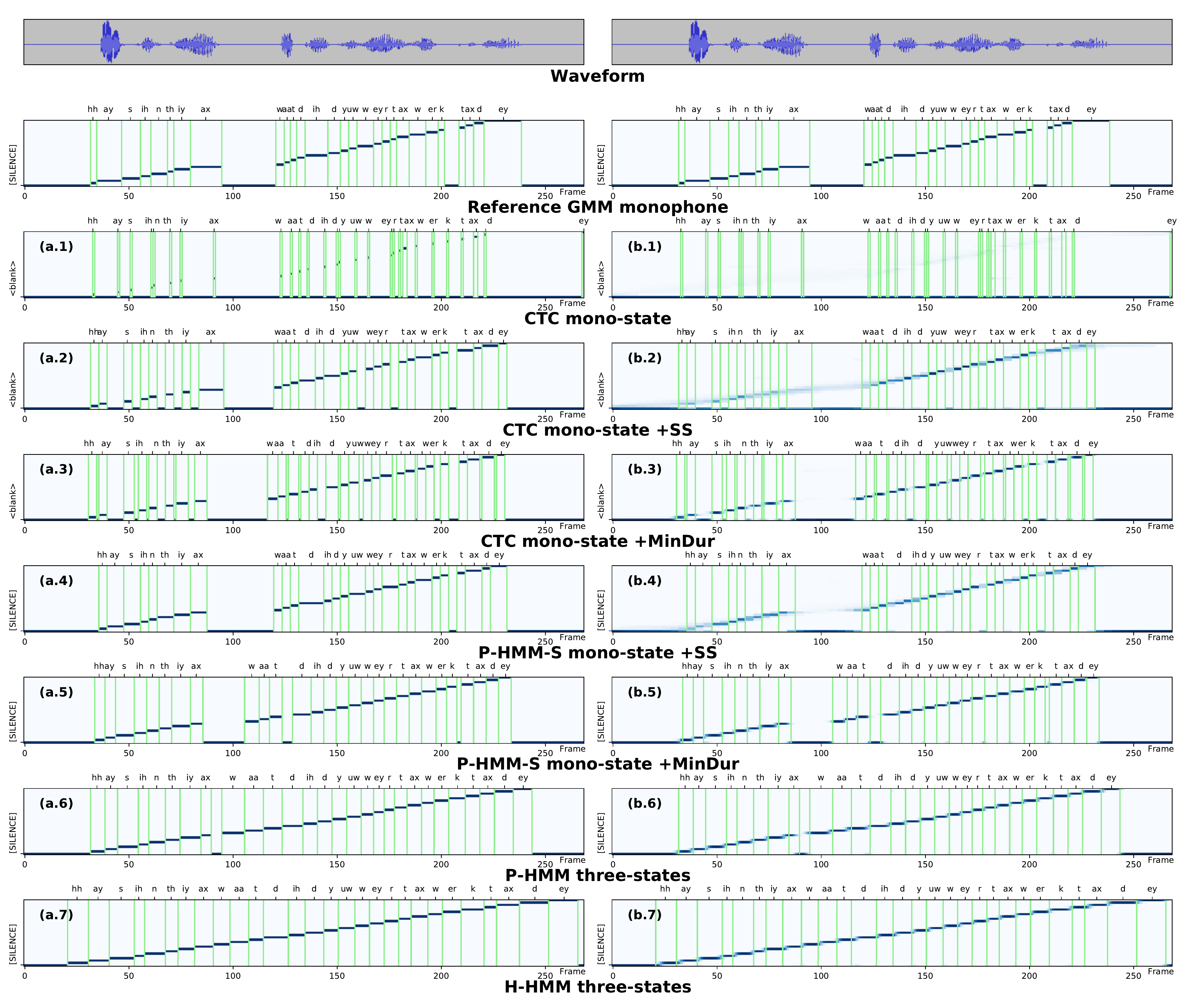}
			\caption{The comparison against a reference GMM monophone alignment of (a) Viterbi and (b) BW paths for different HMM and CTC topologies using our proposed methods for full-sum training from-scratch.\ The horizontal lines indicates the phoneme boundaries.\ }
			\label{fig:fullswbtrain} \vspace{-6mm}
		\end{figure*}
		\subsection{Analysis}
		\label{subsec:analysis}		
		\subsubsection{Alignment Plots}
		\label{subsubsec:plots}
		Whether a high performance ASR model is also able to provide a reasonable alignment with accurate time stamp for subsequent ASR or segmentation tasks is still an open question.\ Therefore, in addition to the evaluation of the ASR accuracy of our models reported in \cref{tab:SWB-WER,tab:LBS-WER}, we carried out further investigations on the quality of the alignment taken from our most promising models.\ For this purpose, we used SWB 300h train and a subset of Hub5'00, which we denote as \textbf{dev-sub}, where all segments containing out-of-vocabulary words with respect to the official SWB lexicon are excluded.\ 
		
		Since the occupation probability is computed using the model parameters at each iteration, we can assume that a soft-alignment with concentrated probability mass around the most prominent path can indicate good model convergence.\ This path eventually is very near or almost the same as what we obtain by doing forced alignment.\ This supposition is confirmed by the plots (a.n) and (b.n) with n $ \in \{1, \cdots, 7\}$, for Viterbi and BW alignments, respectively, as shown in \cref{fig:fullswbtrain}.\ 
		Generally speaking, we believe that the evaluation of the quality of an alignment is yet an undecidable issue.\ There is no ground-truth and therefore no defined evaluation metric for measuring the quality of an alignment.\ This becomes even a more important problem when we compare the alignment across different topologies.\ We chose as reference a GMM monophone alignment, and evaluated our Viterbi alignments by considering the mean absolute distance (in milliseconds) of word start and end-positions against the reference alignment.\ Averaging such distance over all words gives us the time-stamp-error (TSE) metric, as used in~\cite{boostmmi}.\ Note that silence is not counted for TSE, which also makes the evaluation straightforward for CTC models.\ An overview of different TSE values is shown in \cref{tab:tse}.\ Since the alignment learned by a BLSTM base AM can have further shifting due to the NN encoder, we also considered the TSE of a standard H-HMM trained on a tandem system alignment with Fw-CE.\
		
		In \cref{subsec:res}, we showed that subsampling helps to counteract the convergence problem.\ The effectiveness of this method is confirmed also for the alignment quality, where CTC with subsampling has a better TSE.\ This is the same number of 10 ms frames on both train and dev-sub of our P-HMM-S trained with subsampling.\ Even though the ASR accuracy of CTC with minimum duration is worse than CTC with subsampling, we can see that the former produces a better alignment.\ An aspect that is valid also for the case of P-HMM and H-HMM.\ When we switch to the HMM topology, both minimum duration and the choice of the three-states model introduce larger constraint on the phoneme duration, due to the presence of the additional states in the FSA.\ This causes a reduction in the silence duration in the alignment which leads to a larger shift and therefore higher TSE.\ This can be seen when one compares the plots a.4 and a.5 of P-HMM-S with subsampling and minimum duration, respectively.\ The duration of the silence is further reduced in the alignment when we subtract the state prior during training, as a result of a higher penalty on silence.\ This can be seen by comparing the plot a.6 of P-HMM with the plot a.7 of H-HMM.\ Similarly to the CTC case, we have a three-states H-HMM that produces an alignment with higher TSE but has a better ASR accuracy compared to a three-states P-HMM.\ The high silence probability in the latter may force the time synchronous beam search process to wrongly prune away correct speech states.\ However, the absence of the silence in the alignment taken from the H-HMM has the opposite effect of causing a large shift with respect to the evidence in the signal.\ 
		
		\subsubsection{Effect of the Scales for HMM}
		\label{subsubsec:scales}
		The modeling approaches described in \cref{subsec:hmm} make use of optional scales during both training and decoding.\ This means one can control the contribution of each factor via the scales with consequent variations of ASR accuracy.\ We observed that for a given set of fixed scales used during full-sum training, the effect of tuning of the decoding scales on the WER fluctuations are small, similarly to our common experience with Viterbi trained models.\ However, the choice of scales for the full-sum from-scratch training of HMM topology, as shown in \cref{tab:scales}, turns out to be of a significant importance.\  \vspace{-2mm}
		
		\section{Conclusion}
		\label{subsec:conclusion}
		In this work, we performed an in-depth study on from-scratch full-sum training for both CTC and HMM topologies, where we proposed the posterior HMM as a middle ground between standard CTC and hybrid HMM.\ We discussed different methods to counteract the convergence issue during from-scratch full-sum training.\ We showed that by addressing the problem from point of view of difficulty of the alignment modeling, we can improve not only the ASR accuracy, but also the time-stamp-error of our models with respect to a reference GMM alignment.\ Our proposed methods, along with commonly used subsampling, include also novel use of minimum duration and prior-knowledge-based probability approximations.\ Our single-stage trained models show competitive word error rates on Switchboard and LibriSpeech tasks compared to other approaches in the literature.\ For both topologies, we also showed that higher ASR accuracy does not always lead to a better alignment quality.\ We foresee the investigation of a modeling approach that fulfills both requirements as the main future direction of our work.
		
		\section{ACKNOWLEDGMENTS}	
		\footnotesize
			This work was partially supported by NeuroSys which, as part of the initiative “Clusters4Future”, is funded by the Federal Ministry of Education and Research BMBF (03ZU1106DA),  the Deutsche Forschungsgemeinschaft (DFG; grant agreement SCHL2043/2-1, project "Automatic Transcription of Conversations"), and the project HYKIST funded by the German Federal Ministry of Health on the basis of a decision of the German Federal Parliament (Bundestag) under funding ID ZMVI1-2520DAT04A.\ The work reflects only the authors' views and none of the funding parties is responsible for any use that may be made of the information it contains.
	
		\normalsize \newpage
		\bibliographystyle{IEEEbib}
		\bibliography{refs}

	\end{document}